\documentclass{elsart}
\usepackage{graphicx}
\usepackage{amssymb}
\usepackage{amsfonts}
\usepackage{amsmath}
\begin{document}

\begin{frontmatter}
\title{Measurement of the angular distribution in \\
$\bar p p\to \psi(2S)\to e^+e^-$}
\author[FE]{M.~Ambrogiani},
\author[FE]{M.~Andreotti},
\author[TO]{S.~Argiro},
\author[TO,GE]{S.~Bagnasco},
\author[FE]{W.~Baldini},
\author[FE]{D.~Bettoni},
\author[TO]{G.~Borreani},
\author[GE]{A.~Buzzo},
\author[FE]{R.~Calabrese},
\author[TO]{R.~Cester},
\author[FE]{G.~Cibinetto},
\author[FE]{P.~Dalpiaz},
\author[NW]{X.~Fan},
\author[GE,FNAL]{G.~Garzoglio},
\author[UCI,FNAL]{K.~E.~Gollwitzer},
\author[MN]{M.~Graham},
\author[FNAL]{A.~Hahn},
\author[FNAL]{M.~Hu},
\author[NW]{S.~Jin},
\author[NW]{D.~Joffe},
\author[NW]{J.~Kasper},
\author[UCI]{G.~Lasio},
\author[GE]{M.~Lo~Vetere},
\author[FE]{E.~Luppi},
\author[NW]{P.~Maas},
\author[GE]{M.~Macr\'\i},
\author[UCI]{M.~Mandelkern},
\author[TO]{F.~Marchetto},
\author[GE]{M.~Marinelli},
\author[FNAL]{W.~Marsh},
\author[FE]{M.~Martini},
\author[TO]{E.~Menichetti},
\author[NW]{Z.~Metreveli},
\author[TO]{R.~Mussa},
\author[FE]{M.~Negrini},
\author[TO,MN]{M.~M.~Obertino},
\author[GE]{M.~Pallavicini},
\author[TO]{N.~Pastrone},
\author[GE]{C.~Patrignani\corauthref{cor}}
\corauth[cor]{Corresponding author}
\ead{patrignani@ge.infn.it},
\author[NW]{T.~K.~Pedlar},
\author[FNAL]{S.~Pordes},
\author[GE]{E.~Robutti},
\author[UCI]{W.~Roethel},
\author[NW]{J.~Rosen},
\author[NW]{P.~Rumerio},
\author[MN]{R.~Rusack},
\author[GE]{A.~Santroni},
\author[UCI]{J.~Schultz},
\author[MN]{S.~H.~Seo},
\author[NW]{K.~K.~Seth},
\author[FE,FNAL]{G.~Stancari},
\author[UCI,FE]{M.~Stancari},
\author[FNAL]{J.~Streets},
\author[NW]{A.~Tomaradze},
\author[NW]{I.~Uman},
\author[MN]{T.~Vidnovic~III},
\author[FNAL]{S.~Werkema},
\author[UCI]{G.~Zioulas},
\author[NW]{P.~Zweber}
\address[FNAL]{Fermi National Accelerator Laboratory, Batavia, Illinois 60510}
\address[FE]{Istituto Nazionale di Fisica Nucleare and University of Ferrara, 44100, Italy}
\address[GE]{Istituto Nazionale di Fisica Nucleare and University of Genoa, 16146, Italy}
\address[TO]{Istituto Nazionale di Fisica Nucleare and University of Torino, 10125, Italy}
\address[NW]{Northwestern University, Evanston, Illinois, 60208}
\address[UCI]{University of California at Irvine, California 92697}
\address[MN]{University of Minnesota, Minneapolis, Minnesota, 55455}
\collab{Fermilab E835 Collaboration}
\date{\today}

\begin{abstract}
We present the first measurement of the angular distribution for the exclusive 
process $\bar p p \to \psi(2S)\rightarrow e^+e^-$ based on a sample of 
6844 events collected by the Fermilab E835 experiment. 
We find that the angular distribution is well described by the expected functional form
$\frac{dN}{d\cos\theta^{*}}$ $\propto$ 1 + $\lambda \cos^{2}\theta^{*}$, where $\theta^{*}$ is the angle
between the antiproton and the electron in the center of mass frame, with 
$\lambda$ = 0.67 $\pm$ 0.15 (stat.) $\pm$ 0.04 (sys.)
The measured value for $\lambda$ implies a small but non zero $\psi(2S)$ helicity 0 formation 
amplitude in $\bar p p$, comparable to what is observed in $J/\psi$ decays to baryon pairs. 
\end{abstract}


\end{frontmatter}

\section{INTRODUCTION}

The angular distribution of final state electrons from the process 
$\bar p p\to \psi(2S)\rightarrow e^+e^-$
can be written as

\begin{equation}
\frac{dN}{d\cos\theta^{*}} \propto 1 + \lambda \cos^{2}\theta^{*}
\end{equation} 

where $\theta^{*}$ is the angle between an electron and the $\overline{p}$ direction in 
the center-of-mass (CM) system. 

The value of the angular distribution parameter, $\lambda$, is determined by 
the $\psi(2S)$ 
helicity formation amplitudes in $\bar p p$ 

\begin{equation}
\left|\frac{C_{0}}{C_{1}}\right| = \sqrt{\frac{1 - \lambda}{1 + \lambda}}
\end{equation}

with the normalization condition $\left|C_{0}\right|^{2}$ + 2$\left|C_{1}\right|^{2}$ = 1. 

In the  limit of infinitely heavy charm mass, the hadron helicity conservation rule 
implies $\lambda=1$~\cite{brodsky} for both $J/\psi$ and $\psi(2S)$ decays to octet baryon anti-baryon
pairs. Small but not negligible deviations from this prediction
are expected based on constituent quark\cite{carimalo,murgia,ping} or 
hadron mass effect\cite{claudson} from  $\mathcal{O}(v^2)$ and higher twist corrections to the
QCD effective lagrangian, while electromagnetic corrections
are expected to be negligible\cite{carimalo}.

There are several measurements of $\lambda$ in J/$\psi$ decays to baryon anti-baryon pairs 
\cite{mark1,dasp,mark2,dm2,bes}. Only an indirect measurement (with large error) based  
on $\psi(2S)\to J/\psi\,X$, has been
reported for  $\lambda$ at the $\psi(2S)$\cite{psiwid}, and the uncertainty in $\psi(2S)$
helicity formation amplitudes is the major source of systematic error on the $\psi(2S)$
branching ratios measured in E760 and E835\cite{psibr}.

The measurement of $\lambda$ at the $\psi(2S)$ presented here is based on
a sample of 6844 fully reconstructed $\bar p p\to \psi(2S)\to e^+e^-$ events
with negligible ($<1.5\%$) background.

\section{E835 DETECTOR}

The detector and the experimental technique are described in detail in 
~\cite{e835_det}. Here
we recall only the features relevant to the present work.

The experiment was located in the Antiproton Accumulator (AA) ring at Fermilab.
The stochastically cooled antiproton beam ($\Delta p/p\approx10^{-4}$) circulating in the
AA passed through an internal hydrogen gas-jet target. The energy of the beam
could be tuned to the charmonium resonance of interest, in this case
the $\psi(2S)$. The E835 detector
was a nonmagnetic spectrometer with cylindrical symmetry about the
beam axis. The inner part of the detector contained a system for precise
tracking of charged particles and four scintillator hodoscopes used
variously for triggering and $dE/dx$ measurement. Outside the inner
detectors was a 16 cell threshold \v{C}erenkov counter to identify electrons
at the trigger level and offline; the \v{C}erenkov covered the polar
angle range from 15$^\circ$ to 65$^\circ$. Two electromagnetic calorimeters
used to measure the angles and energies of photons and electrons
completed the detector. The Central Calorimeter (CCAL), composed of 1280
lead-glass {\v C}erenkov counters arranged in a pointing geometry, covered
the polar angle region from 11$^\circ$ to 70$^\circ$. The energy resolution
of the CCAL was $\sigma(E)/E= 6\%/\sqrt{E}(GeV) + 1.4\%$. Given the
size of the target region ($\approx0.6$~cm$\,\times\,0.6$~cm$\,\times\,0.6$~cm), the angular
resolution for photons and electrons was 6~mrad in polar
angle ($\theta$) and 11~mrad in azimuth ($\phi$).  The Forward Calorimeter which
covered the region from 3$^\circ$ to 11$^\circ$ is not used in this analysis.
The luminosity was measured by a set of solid state detectors which counted
recoil protons from elastic scatters at 90$^\circ$.

The calorimeter channels were equipped with both ADC's to record pulse
height and TDC's to record the time with respect to the trigger. The latter
allowed signals from accidentals within the ADC gate to be ignored thus maintaining analysis
efficiency at high luminosity.
Signals were labelled `in-time' if they were within $\pm$10~ ns of the trigger,
`out-of-time' if they were outside
this range. Channels with no timing information (the TDC threshold for small
energy deposits was $\approx50$~MeV) were labelled `undetermined'.

\section{EVENT SELECTION}

Two sets of $\psi(2S)$ data from different data taking periods (1996-1997 and 2000) 
were used for this analysis.

The total luminosity for these data sets is 22.57~pb$^{-1}$: 10.09 pb$^{-1}$  
for the 1996-1997 run and 12.48 pb$^{-1}$  for the 2000 run. 
The typical instantaneous luminosity during data taking was
$\approx 2\times10^{31}$cm$^{-2}$s$^{-1}$.


The hardware trigger was designed to accept
events with a large-mass $e^+e^-$~pair within the acceptance of the central
calorimeter. It required two ``electron tracks",
defined by the appropriate coincidence of the inner and outer scintillator
hodoscopes and the corresponding cell of the \v{C}erenkov counter, and 
independently two large energy deposits (clusters) in CCAL
separated by $>90^\circ$ in azimuth, with an invariant mass
$>2.2$ GeV.

Offline reconstruction of electron showers in the CCAL was performed clustering all 
hits in a 5$\times$5 grid around a central block ({\it seed}) with at least 50~MeV energy deposit.
If the ``cluster mass'' defined as

\begin{equation} 
M_{cl} \equiv \sqrt{\left(\sum_{i=1}^{N_{blk}}E_{i}\right)^{2} - \left(\sum_{i}\vec{p}_{i}\right)^{2}}
\end{equation}

exceeded $M_{cl}>120$~MeV, the cluster is considered as originating from
two overlapping e.m. showers and it is split into two distinct clusters.
The values for the seed energy (50~MeV) and the cluster splitting mass
(120~MeV) are specific to the $\psi(2S)\to e^+e^-$ channel and were chosen to
ensure reasonably uniform efficiency over the angular acceptance.

A preliminary selection, aimed at generic $e^+e^-\, X$ channels, 
required the two highest energy clusters in the
calorimeter to have $M_{e^+e^-}>$2.6~GeV and to be 
associated to hits in the \v{C}erenkov
and in at least two of the three scintillators.
To reject background, largely due to Dalitz decay or photon conversions of $\pi^0$s misidentified as
single electrons, we calculate the likelihood ratio (EW) of the electron and background
hypothesis (described in detail in \cite{e835_det}) and require that
$EW_1\times EW_2>10^{-4}$.

A four constraint kinematic fit to the hypothesis $\bar p p\to  e^+e^-$ is performed on all
events with two high energy clusters (candidate electron-positron pair).
Events with no extra clusters were retained if
the nominal $\chi^2$ probability $Prob(\chi^2)>10^{-5}$. 
Events with up to two extra clusters (either on-time or undetermined) were also retained if the
$Prob(\chi^2)>10^{-2}$. This was done to retain events where the high energy electron shower 
was not contained within the 5x5 grid or the electron had radiated a bremsstrahlung photon 
in the material in front of the calorimeter.

\begin{figure}
\begin{center}
\includegraphics[scale=0.5]{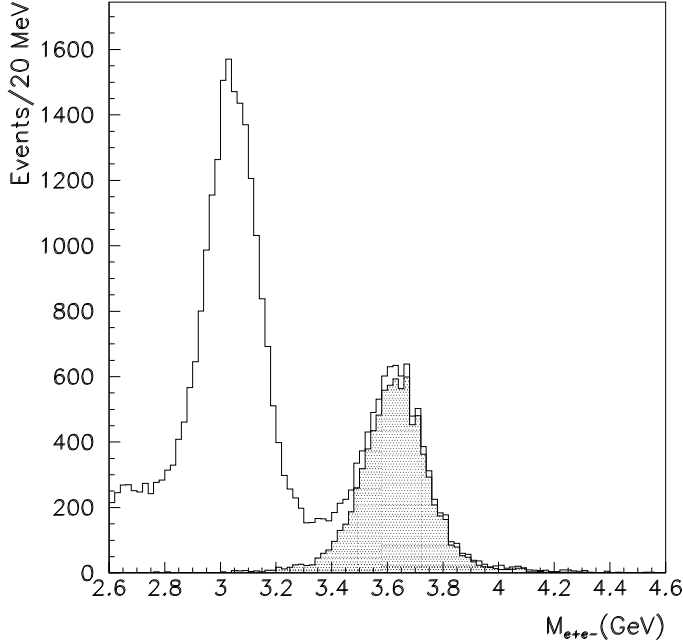} 
\end{center}
\caption{ Invariant mass distributions of $\bar p p\to e^{+}e^{-}\,(X)$ candidates. 
The shaded area represents events in the $\bar p p\to \psi(2S)\to e^{+}e^{-}$ sample.} 
\label{mee}
\end{figure}

To avoid possible contamination from $\psi(2S)\to J/\psi\,X$ events, we finally require that
$M_{e^+e^-}>3.4$~GeV (see fig.~\ref{mee}).

The amount of material in front of the calorimeter
and the size of the calorimeter counters both varied with angle.
This means that the probability of bremsstrahlung and the number
of low energy satellite clusters distinct from the high energy
cluster could vary with angle. 

\begin{figure}
\begin{center}
\includegraphics[scale=0.5]{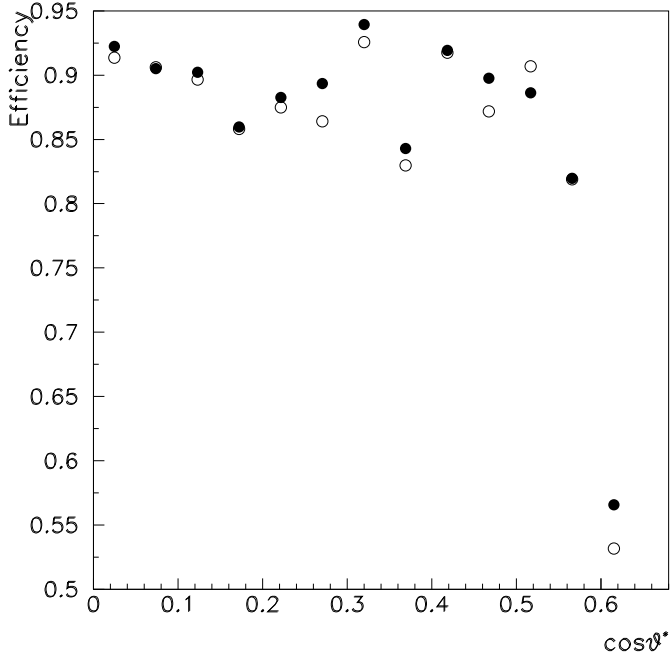}
\end{center}
\caption{ Selection efficiency as a function of $\cos{\theta^{*}}$ for the 1996-1997 
(open circles) and 2000 (filled circles) data sets.}
\label{eff}
\end{figure}

A full Monte Carlo (MC) detector simulation based on GEANT~\cite{geant} 
was performed to evaluate the efficiency correction as a function
of $\cos{\theta^{*}}$ (shown in fig.~\ref{eff}). 

With the cuts chosen, the efficiency is essentially independent of angle.
 
To ensure uniform efficiency we limit our 
acceptance to $\vert\cos{\theta^{*}}\vert<0.58$, where the efficiency
ranges between 0.83 and 0.93. The
bin by bin differences depend mainly on the CCAL counters' 
geometry (including two dead channels) that are accurately modelled in the MC.

The statistical error on the efficiency
due to the size of the MC sample (330,000 events)
is negligible given the size of our data sample.

The final sample has 2391 and 4457 events from the  1996-1997 and 2000 runs, respectively.

There are two possible sources of background events, the first from genuine electrons and 
positrons
from 
$\psi(2S)\to J/\psi\,X$ events mis-classified as exclusive $\psi(2S)\to e^+e^-$ decays,
the second from events where the electron and positron candidates are 
from non resonant hadronic events with $\pi^0$ Dalitz decays or photon conversions in the
beam pipe.

The backgrounds from mis-classified events is estimated performing the same analysis
on Monte Carlo samples of 100,000 $\psi(2S)$ decays to J/$\psi$ $\eta$, J/$\psi$ $\pi^{0}$$\pi^{0}$ and 
J/$\psi$ $\pi^{+}$ $\pi^{-}$. We expect less than 1~\% contamination and no subtraction is 
performed. 

The background from mis-identified electron-positron pairs is measured using samples of data taken off-resonance,
at center of mass energies  3576 MeV $<$ $\sqrt{s}$ $<$ 3660 MeV in 1996-1997 and 
$\sqrt{s}$=3666, 3705 and 3526 MeV in 2000.
The mis-identified background contamination is less than 0.4~\% in both runs,
and also in this case no background subtraction is performed.

\section{RESULTS}


Binned likelihood fits were performed on the 1996-1997 and the 2000 data sets separately.
Data were binned in bins of 2.1$^{o}$, 
corresponding to the average CCAL block polar coverage (1.52$^{o}$ $\sim$ 4.80$^{o}$). 

The angular dependence of the efficiency correction was taken into account on a bin by bin basis.

\begin{figure}
\centerline{\includegraphics[scale=0.4]{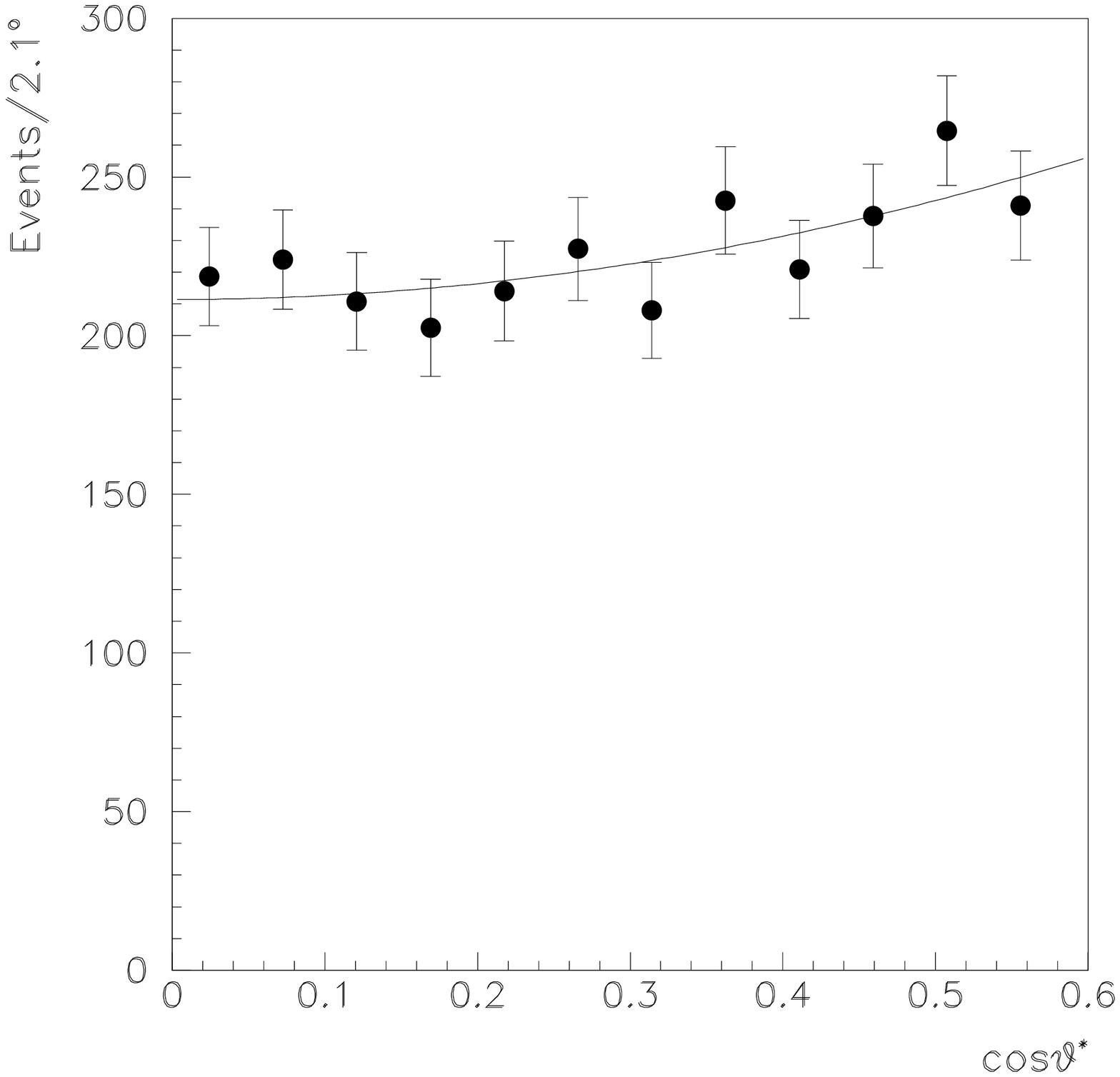}
\includegraphics[scale=0.4]{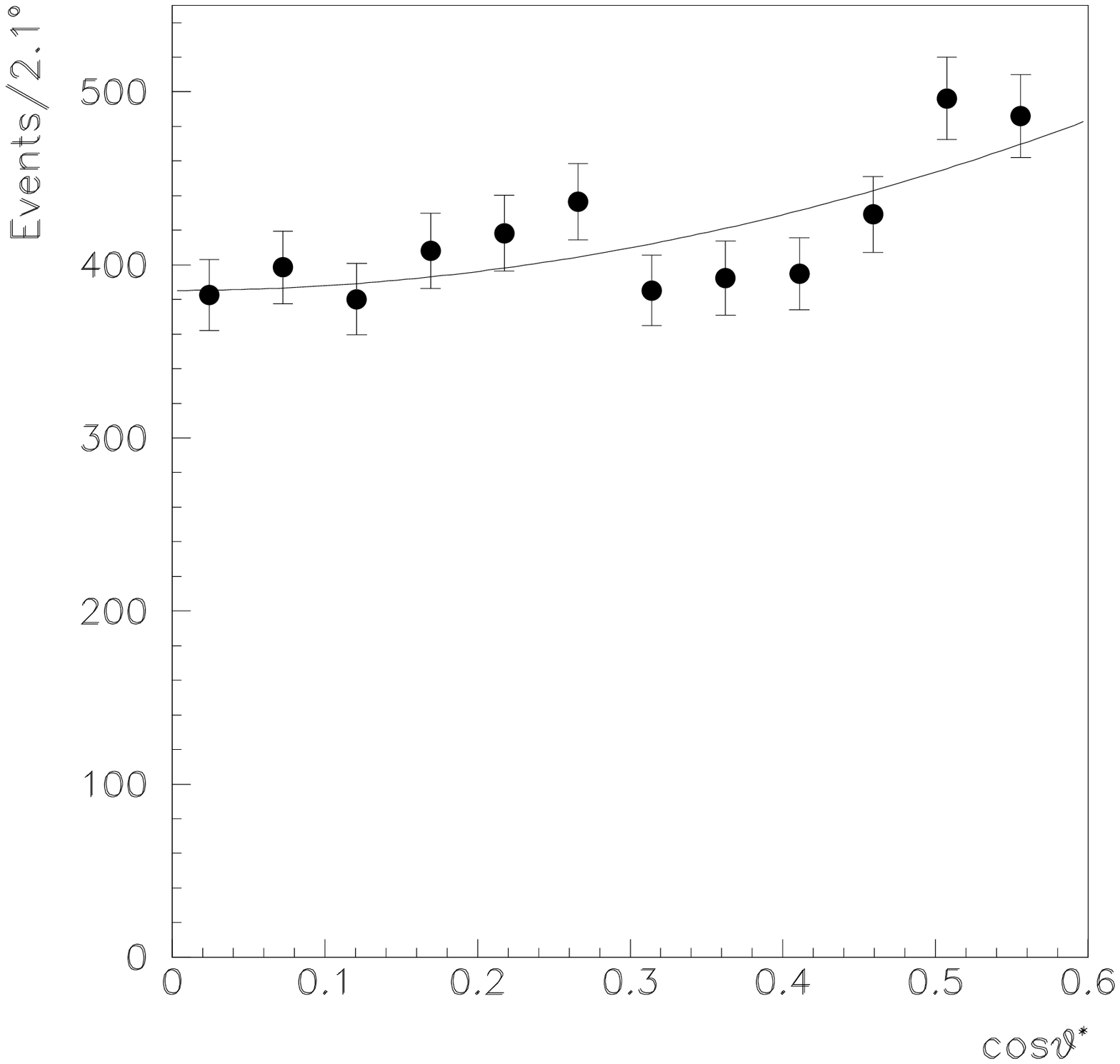}}
\caption{Angular distributions after efficiency correction for the 1996-1997 (left) and 
2000 data (right). The lines represent the likelihood fit to data.}
\label{each}
\end{figure}

Results are summarized in table~\ref{results} and shown in  
fig.~\ref{each} for each data set.

The systematic errors were estimated by varying the cluster seed threshold and
the cluster mass value used in electron shower reconstruction, 
the kinematic fit $\chi^2$ probability, the EW cut, and the invariant mass cut
used in event selection. The systematic error, expected to be common to both data sets,
has been estimated separately on the two samples for each of the above sources 
to verify this assumption. No significant correlation was found between the systematics 
from different sources, and the total systematic error (0.04) has been evaluated adding in quadrature 
the contribution from all sources.

The bin by bin differences in efficiency are smaller than the statistical 
uncertainty of data in each bin, and
the contribution to the systematic error due to the uncertainty in the efficiency 
correction is negligible compared to other sources.

Further details on the analysis and on the systematic error evaluation can be found in \cite{seon-hee}.

\begin{table}
\begin{centering}
\begin{tabular}{|c||c|c|}
\hline
  &  { 1996-1997 Data} & { 2000 Data}  \\
\hline
Candidate events & 2391 & 4453 \\
(0$<\cos\theta^{*}<$0.58) & & \\
\hline
$\lambda$ & 0.59 $\pm$ 0.24  & 0.71 $\pm$ 0.18 \\
\hline
\multicolumn{3}{|c|}{Sources of systematic error}\\
\hline
Cluster seed threshold & \multicolumn{2}{c|}{$\pm$0.01}\\
$Prob(\chi^2)$ & \multicolumn{2}{c|}{$\pm$0.01} \\
EW cut               & \multicolumn{2}{c|}{ --} \\
Cluster mass        & \multicolumn{2}{c|}{$\pm$0.01}\\
$M_{e^+e^-}$        & \multicolumn{2}{c|}{$\pm$0.04} \\
\hline
Total systematic  & \multicolumn{2}{c|}{$\pm$0.04}  \\
\hline
\end{tabular}
\caption{ Results for the two data sets.}
\label{results}
\end{centering}
\end{table}

While the CCAL and \v{C}erenkov remained the same for both sets of data,
the amount of material in the inner detectors was about
half as much for the 2000 data as for the 1996-1997 data, resulting 
in a small difference 
in the angle dependent efficiency correction (see fig.~\ref{eff}) between
the two data sets.

\begin{figure}
\centerline{\includegraphics[scale=0.5]{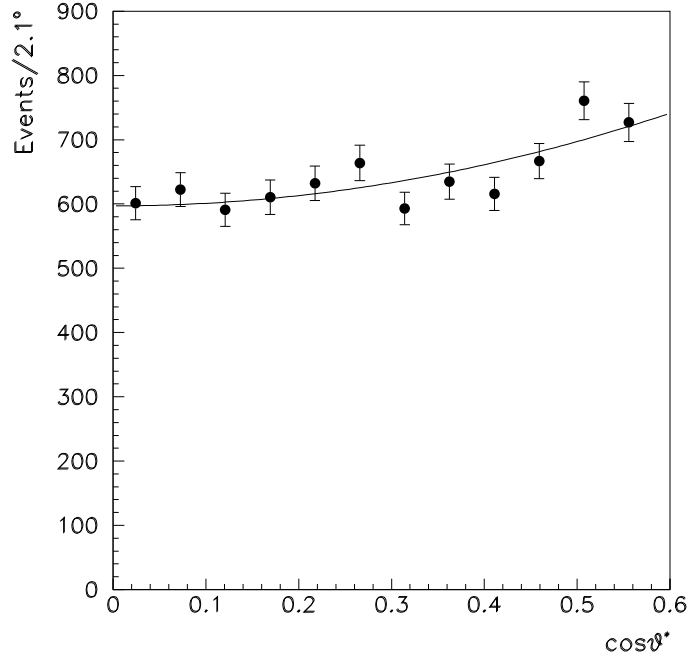}}
\caption{Angular distribution from the combined data set. 
The line represents the  fit result $\lambda$ = 0.67.}
\label{comb}
\end{figure}

Based on the Kolmogorov-Smirnov test ~\cite{ks} the probability for the $\cos{\theta^\ast}$ 
distributions measured in the two periods of data taking to be compatible with the same
angular distribution is 74.4$\%$. We therefore perform the likelihood fit to the combined
data sets (shown in fig.~\ref{comb} ) and obtain $\lambda$ = 0.67 $\pm$ 0.15(stat.)$\pm$0.04 (sys.). 

The corresponding ratio of the  $\psi(2S)$ helicity formation amplitudes is 
 $\left|\frac{C_{0}}{C_{1}}\right|_{\psi(2S)} =  0.44 \pm 0.12\pm0.03$.

\section{CONCLUSIONS}

We have presented the first measurement of the angular distribution parameter $\lambda$
at the $\psi(2S)$. 

From this measurement we determine that the helicity 
amplitude ratio at the $\psi(2S)$ is
$$\left|\frac{C_{0}}{C_{1}}\right|_{\psi(2S)} =  0.44 \pm 0.12\pm0.03.$$

\begin{table}
  \begin{center}
  \begin{tabular}{|l|c|c|} \hline
  Reference  & $\lambda_{(J/\psi)}$ &  $\lambda_{(\psi(2S))}$ \\
\hline
\multicolumn{3}{|c|}{Predicted}\\
\hline
Claudson et al.\cite{claudson} ~(eq. 9)                   & 0.46 & 0.58\\   
Carimalo\cite{carimalo}~(eq. 24)                    & 0.69 & 0.80 \\     
\hline
	\multicolumn{3}{|c|}{Measured} \\
\hline
MARK-II\cite{mark2}& $0.61 \pm 0.23$ & -- \\
DM2\cite{dm2}  & $0.62\pm0.11$ & -- \\
BES\cite{bes}  & $ 0.676\pm0.036\pm 0.042 $ & -- \\
    This experiment &  -- & $0.67\pm0.15\pm0.04$ \\
\hline 
  \end{tabular}
  \end{center}
  \caption[]{Experimental results and theoretical predictions for the parameter $\lambda$
    in $J/\psi,\psi(2S)\to p\bar p$.}
\label{lambdata}
\end{table}

The value of $\lambda$ measured at the J/$\psi$ (table~\ref{lambdata}) is 0.66~$\pm$~0.05 , 
which results in $\left|\frac{C_{0}}{C_{1}}\right|_{J/\psi} =  0.45 \pm 0.04$.
 
The ratio of the helicity amplitudes is the same within the errors at the J/$\psi$ 
and $\psi(2S)$.

\section{ACKNOWLEDGMENTS}

The authors wish to thank the staffs, engineers and technicians at 
their respective institutions for their valuable help and cooperation. 
This research was supported by the U.S. Department of Energy and by the 
Italian Istituto Nazionale di Fisica Nucleare. 


\end{document}